\documentclass[aps,bibnotes,twocolumn,showpacs]{revtex4}
\usepackage[tbtags]{amsmath}
\usepackage{bm}

\def\beq{\begin{equation}}
\def\eeq{\end{equation}}
\def\eeql#1{\label{#1} \end{equation}}
\def\C{\Gamma}
\def\e{\epsilon}
\def\w{\omega}
\def\hn{\mskip-0.5\thinmuskip}
\def\hp{\mskip0.5\thinmuskip}
\newcommand{\To}{\rightarrow}
\newcommand{\rs}{r_{\!*}}
\newcommand{\dsf}{{\displaystyle\frac{y}{2}}}
\newcommand{\ts}{\textstyle}
\newcommand{\lb}{\mathopen\ll}
\newcommand{\php}{{\vphantom{\prime}}}
\newcommand{\rb}{\mathclose\gg}
\newcommand{\Oc}{\mathcal{O}}
\DeclareMathOperator{\sh}{sh}
\DeclareMathOperator{\ch}{ch}
\DeclareMathOperator{\re}{Re}

\begin{document}

\title{Approach to the extremal limit of the Schwarzschild--de Sitter black hole}

\author{Alec \surname{Maassen van den Brink}}
\email{alec@dwavesys.com}
\affiliation{Physics Department, The Chinese University of Hong Kong, Hong Kong, China}
\date{\today}


\begin{abstract}
The quasinormal-mode spectrum of the Schwarzschild--de Sitter black hole is studied in the limit of near-equal black-hole and cosmological radii. It is found that the mode \emph{frequencies} agree with the P\"oschl--Teller approximation to one more order than previously realized, even though the effective \emph{potential} does not. Whether the spectrum approaches the limiting one uniformly in the mode index is seen to depend on the chosen units (to the order investigated). A perturbation framework is set up, in which these issues can be studied to higher order in future.
\end{abstract}


\maketitle

Since Schwarzschild's original black-hole solution of the Einstein equations, several extensions have been found, including those with nonzero rotation or electric charge~\cite{chand}, or the Schwarzschild--de Sitter (SdS) hole with a finite outer `cosmological' horizon~\cite{SdS}. Each of these leads to a corresponding generalization of the wave equation governing the linearized perturbations~\cite{R-W}. Relaxing the requirement of physical significance, one can also generalize this wave equation by staying with the Schwarzschild spacetime, but taking the spin $s$ of the perturbing field to be formally continuous (gravity waves correspond to $s=2$). In all these cases, one faces the question how the Schwarzschild quasinormal-mode (QNM) spectrum is modified. This question is complicated by said spectrum being infinite, with two strings of high-damping modes---currently undergoing intense scrutiny~\cite{hod}---extending to $\w=-i\infty$; therefore, one can have nontrivial interplay between the Schwarzschild and large-mode-index limits~\cite{neitzke}.

Clearly, the study of these and analogous situations yearns for a solvable zeroth-order model. Recently, one has become available by the demonstration~\cite{Cardoso} that the SdS wave equations reduce to the familiar P\"oschl--Teller (PT) model~\cite{PT} in the \emph{opposite} `extremal' limit. While the PT potential was already being used as a toy model for gravitational problems~\cite{Ferrari}, the derivation puts this on a firm footing and in particular provides a class of physically motivated perturbations, for which moreover numerical data are already available~\cite{MN}. In this report, these perturbing potentials are derived in detail, and a beginning is made with their study.

The SdS spacetime is characterized by a black-hole horizon, set to unity here, and a cosmological horizon~$c$, the only nontrivial parameter. For each angular momentum $\ell$, the perturbations obey a wave equation
\beq
  [d_{\rs}^2-V\bm{(}r(\rs)\bm{)}+\w^2]\hp\psi(\rs,\w)=0\;,
\eeql{WE}
where the tortoise coordinate is given by $d\rs=dr/f(r)$ in terms of the warp factor
\beq
  f=\frac{(r-1)(c-r)(r+1+c)}{rA}\;,
\eeql{f}
with $A\equiv1+c+c^2$. Depending on the type of perturbation (i.e., on~$s$), the effective potential reads~\cite{Cardoso}
\beq
  \frac{V}{f}=\left\{\negthickspace\begin{array}{ll}
  \ell(\ell{+}1)r^{-2}+2Mr^{-3}-2/A\;, & \text{scalar}\;; \\
  \ell(\ell{+}1)r^{-2}\;, & \text{electromagnetic}\;; \\
  \ell(\ell{+}1)r^{-2}-6Mr^{-3}\;, & \text{gravitational}\;,\end{array}\right.
\eeql{V}
with $M=c(1{+}c)/2A$. In Ref.~\cite{MN}, the ensuing QNM spectra are compared to a PT approximation for all~$c$. However, the used fit amounts to reproducing the \emph{fundamental} (complex) frequency to high accuracy, while in fact it is the \emph{high-order} modes that---at least in the Schwarzschild limit---qualitatively have PT character. Thus, there is room for further analytical study of the PT approximation.

For $c\To\infty$, (\ref{WE})--(\ref{V}) readily reduce to the familiar Regge--Wheeler equation for the Schwarzschild spacetime. Here, however, we are interested in the limit $c\downarrow1$, where the inner and outer radii merge. Let us scale $c=1+\e$, $r=1+\e\xi$ ($0<\xi<1$), $\rs=x/\e$, $V=\e^2W$, and $\w=\e\nu$. The new tortoise coordinate $x=\e^2\!\int\!d\xi/f$ is
\beq\begin{split}
  x=(3{+}3\e{+}\e^2)&\biggl[\frac{\ln\xi}{3{+}\e}
  -\frac{1{+}\e}{3{+}2\e}\ln(1{-}\xi)\\ &\hphantom{\biggl[}
  +\frac{\e(2{+}\e)}{(3{+}\e)(3{+}2\e)}\ln(3{+}\e{+}\e\xi)\biggr]\;.
\end{split}\eeql{x-xi}
In the scaled variables, expansion in $\e$ is straightforward. Inverting $x(\xi)$, one finds
\beq
  \frac{\xi}{\xi_0}=1-\frac{2\e}{3}\frac{\ln(1+e^x)+x}{1+e^x}+\Oc(\e^2)\;,
\eeql{xi-x}
with $\xi_0\equiv1/(1{+}e^{-x})$. Although we intend to carry our expansions to second order, we will only do so after one further simplification. Namely, consider electromagnetic perturbations. Inserting (\ref{xi-x}) into (\ref{V}), one obtains
\beq
  \frac{W_\mathrm{el}}{\ell(\ell{+}1)}\approx
  \frac{1{-}2\e}{4\ch^{\hn2}\hn(x/2)}+
  \frac{\e}{6}\Bigl[\ln\Bigl(2\ch\!\frac{x}{2}\Bigr)+\frac{3x}{2}-2\Bigr]
  \frac{\sh(x/2)}{\ch^{\hn3}\hn(x/2)}\,.
\eeql{Wel}
One observes that most of the $\Oc(\e)$-corrections can in fact be absorbed into a further redefinition of the coordinate. In other words, it is---unsurprisingly---advantageous to fit $V_\mathrm{el}$ to a PT potential with $\e$-dependent position and width. Thus, we set $x=(1{+}\e)y-\frac{4}{3}\e$, $W=U/(1{+}\e)^2$, and $\nu=\mu/(1{+}\e)$, leading to [$\xi_1\equiv1/(1{+}e^{-y})$]
\begin{widetext}
\begin{gather}
  \frac{\xi}{\xi_1}=1-\frac{2\e}{3}
  \frac{\ln\Bigl(2\ch\!\dsf\Bigr)+2}{1+e^y}
  +\frac{\e^2}{9}\frac{2(1{-}e^y)\ln^2\!\Bigl(2\ch\!\dsf\Bigr)
  +(9{-}3e^y)\ln\Bigl(2\ch\!\dsf\Bigr)
  -\frac{5}{2}(1{+}e^y)y+16+6e^y}{(1+e^y)^2}+\Oc(\e^3)\;,\displaybreak[0]\\
  \begin{aligned}[b]\frac{U_\mathrm{el}}{\ell(\ell{+}1)}&=
  \frac{1}{4\ch^{\hn2}\hn(y/2)}
  +\frac{\e}{6}\frac{\ln\Bigl(2\ch\!\dsf\Bigr)\sh\!\dsf}{\ch^{\hn3}\hn(y/2)}\\
  &\quad+\frac{\e^2}{144}\frac{2(e^y{-}4{+}e^{-y})\ln^2\!\Bigl(2\ch\!\dsf\Bigr)
  +(4{-}5e^y{+}e^{-y})\ln\Bigl(2\ch\!\dsf\Bigr)+5\sh(y)y+5-10e^y}
  {\ch^{\hn4}\hn(y/2)}+\Oc(\e^3)\;.\end{aligned}\label{Uel}
\end{gather}
Since the $\Oc(\e)$-term is now \emph{odd}~\cite{odd}, in these coordinates there are no $\Oc(\e)$-corrections at all. However, this $\Oc(\e)$-term in $U_\mathrm{el}$ can contribute in $\Oc(\e^2)$, see the Appendix. For studying perturbation to the latter order, the $\Oc(\e^2)$-term in $U_\mathrm{el}$ can be safely symmetrized, so that the numerator becomes
\beq
  \Bigl[8\ch^{\hn2}\hn\Bigl(\dsf\Bigr){-}12\Bigr]\ln^2\!\Bigl(2\ch\!\dsf\Bigr)
  -8\sh^{\hn2}\hn\Bigl(\dsf\Bigr)\ln\Bigl(2\ch\!\dsf\Bigr)
  +10\ch\Bigl(\dsf\Bigr)\sh\Bigl(\dsf\Bigr)y
  -20\ch^{\hn2}\hn\Bigl(\dsf\Bigr)+15\;.
\eeq

For scalar and gravitational perturbations, the analogous potentials read
\begin{align}
  U_\mathrm{s}&=U_\mathrm{el}-\frac{\e}{4}\frac{\sh(y/2)}{\ch^{\hn3}\hn(y/2)}
    +\frac{\e^2}{48}\frac{2(4{-}e^y{-}e^{-y})\ln\Bigl(2\ch\!\dsf\Bigr)
    +e^y+7-2e^{-y}}{\ch^{\hn4}\hn(y/2)}+\Oc(\e^3)\;,
    \label{Us}\displaybreak[0]\\[2mm]
  U_\mathrm{gr}&=U_\mathrm{el}-\frac{2}{4\ch^{\hn2}\hn(y/2)}
    +\frac{\e}{12}\Bigl[3-4\ln\Bigl(2\ch\!\dsf\Bigr)\Bigr]
    \frac{\sh(y/2)}{\ch^{\hn3}\hn(y/2)}\notag\\
  &\quad+\frac{\e^2}{144}\frac{4(4{-}e^y{-}e^{-y})\ln^2\!\Bigl(2\ch\!\dsf\Bigr)
    +(16e^y{-}32{+}4e^{-y})\ln\Bigl(2\ch\!\dsf\Bigr)-10\sh(y)y+17e^y-13+6e^{-y}}
    {\ch^{\hn4}\hn(y/2)}+\Oc(\e^3)\;.\label{Ugr}
\end{align}
\end{widetext}
Again, the leading corrections are odd. The $\e\sh(y/2)\hn\*\ch^{\hn-3}\hn(y/2)$-terms could be eliminated also here by making the $x\mapsto y$ shift $s$- and $\ell$\nobreakdash-dependent, but this will not be pursued. Only after the first-order terms in (\ref{Us})--(\ref{Ugr}) will thus have been maximally simplified does it look meaningful to symmetrize the second-order terms.

Defining
\beq
  q=\left\{\negthickspace\begin{array}{ll}
  \sqrt{\ell(\ell{+}1)-\frac{1}{4}}\;,& \text{s, el}\;;\\[2mm]
  \sqrt{\ell(\ell{+}1)-\frac{9}{4}}\;,& \text{gr}\;,\end{array}\right.
\eeq
in all cases one has $\mu_n=\frac{1}{2}[q-(n{+}\frac{1}{2})i]+\Oc(\e^2)$, $n=0,1,2,\ldots$, so that $\w_n={\ts\frac{1}{2}}(\e{-}\e^2)[q-(n{+}{\ts\frac{1}{2}})i]+\Oc(\e^3)$. An important quantity in SdS is the surface gravity~\cite{Cardoso}
\beq
  \kappa=\frac{\e(3{+}\e)}{2(3{+}3\e{+}\e^2)}
  =\frac{\e}{2}-\frac{\e^2}{3}+\frac{\e^3}{6}+\cdots
\eeql{kap}
Since $0<\e<\infty$ but $0<\kappa\le\frac{1}{2}$, it is heuristically advantageous to reparametrize the problem in terms of~$\kappa$, inverting (\ref{kap}) as $\e=2\kappa+\frac{8}{3}\kappa^2+\frac{40}{9}\kappa^3+\cdots$, so that
\beq
  \w_n=\kappa(1{-}{\ts\frac{2}{3}}\kappa)[q-(n{+}{\ts\frac{1}{2}})i]
  +\Oc(\kappa^3)\;.
\eeql{res}
The factor $1-\frac{2}{3}\kappa$ is also the leading one in (19) of~\cite{MN}, but here the general $\ell$-dependence has been obtained exactly. For a merely moderately small $\kappa=0.01411708$, (\ref{res}) yields
\beq
  \w_{0,\mathrm{gr}}\approx\left\{\negthickspace\begin{array}{ll}
    0.0270803-0.00699211i\;, & \quad\ell=2\;; \\
    0.0436657-0.00699211i\;, & \quad\ell=3\;, \end{array}\right.
\eeq
while numerically one has~\cite{MN}
\beq
  \w_{0,\mathrm{gr}}=\left\{\negthickspace\begin{array}{ll}
    0.0270837-0.00699266i\;, & \quad\ell=2\;; \\
    0.0436710-0.00699280i\;, & \quad\ell=3\;. \end{array}\right.
\eeq
For both values of $\ell$, the error has been verified to decrease cubically with $\kappa$.

Thus, $\w_n/\e$ does have $\Oc(\e n)$-corrections [$\w_n/\kappa$ has $\Oc(\kappa n)$-corrections]. In these units one could say that there is `non-uniform convergence to the limit', while in fact $\mu_n$ does not have any corrections at all in this order. Of course, it is possible (even likely) that the $\Oc(\e^2)$-corrections to $\mu_n$ are unbounded in~$n$. The above, plus the general perturbation theory in the Appendix, provide the framework for studying the issue further. However, the relevant (regularized) integrals are unlikely to yield to exact evaluation and thus will have to be studied asymptotically in~$n$, which is beyond the present scope.


\begin{acknowledgments}
I thank V. Cardoso for correspondence stimulating this paper, J.P. Norman for sharing his data, the CUHK group for previous collaboration on the P\"oschl--Teller model, and H.Y. Guo for supporting my calculations.
\end{acknowledgments}

\appendix

\section{P\"oschl--Teller potential}

The PT QNMs are well known~\cite{Ferrari}, but the present recapitulation adds a normalization for the eigenfunctions, as it applies to the perturbation theory which is also sketched. Consider
\begin{gather}
  [d_x^2-V(x)+\w^2]\hp\psi(x,\w)=0\;,\label{we-App}\\
  V(x)=\frac{V_0}{\ch^{\hn2}\hn(x/2)}\;,\label{V-pt}
\end{gather}
and define $q=\sqrt{4V_0-\frac{1}{4}}$. The outgoing wave to the left, $f(x{\To}{-}\infty)\linebreak[0]\sim1\cdot e^{-i\w x}$, reads
\beq\begin{split}
  f(x,\w)&=[\xi(1{-}\xi)]^{-i\w}\\ &\quad\times
  F({\ts\frac{1}{2}}{+}iq{-}2i\w,{\ts\frac{1}{2}}{-}iq{-}2i\w;1{-}2i\w;\xi)\;,
\end{split}\eeql{f-pt}
where $F$ is the hypergeometric function and $\xi\equiv1/(1{+}e^{-x})$. The outgoing wave to the right, $g(x{\To}\infty)\linebreak[0]\sim1\cdot e^{i\w x}$, follows as $g(x,\w)=f(-x,\w)$, i.e., $\xi\mapsto1-\xi$ in~(\ref{f-pt}). Using the connection formulas for $F$ yields
\beq\begin{split}
  g(x,\w)&=\frac{\C(1{-}2i\w)\hp\C(-2i\w)}
  {\C(\frac{1}{2}{+}iq{-}2i\w)\hp\C(\frac{1}{2}{-}iq{-}2i\w)}f(x,-\w)\\[1mm]
  &\quad+\frac{\sin[\pi(\frac{1}{2}{+}iq)]}{\sin[2\pi i\w]}f(x,\w)\;.
\end{split}\eeq
The Wronskian $J(\w)=gf'-fg'$ becomes
\beq
  J(\w)=\frac{\C(1{-}2i\w)^2}
    {\C(\frac{1}{2}{+}iq{-}2i\w)\hp\C(\frac{1}{2}{-}iq{-}2i\w)}\;,
\eeq
and has zeros $2\w_n=\pm q-(n{+}\frac{1}{2})i$, $n=0,1,2,\ldots$ Here, we focus on real~$q$ and, without loss of generality, on the QNMs in the fourth quadrant (plus sign), as has tacitly been done in the main text. The derivative is
\begin{align}
  J'(\w_n)&=(-)^{n+1}2in!\hp\frac{\C(\frac{1}{2}{-}n{-}iq)^2}{\C(-n{-}2iq)}\\
  &=-2\w_n\lb f_ng_n\rb\;,
\end{align}
where the second line is a general relation between the Wronskian and the norm of the eigenfunctions~\cite{rmp}. For potentials decaying as a series of exponentials [like (\ref{V-pt})], this norm $\lb a\rb$ is defined as the constant term in the asymptotic expansion of $\int_{x_1}^{x_2}\!dx\,a(x)$ for $-x_1,x_2\To\infty$. Since the parity of $V$ presently implies that $g_n=(-)^nf_n$,
\begin{gather}
  \lb f_n^2\rb=2n!\hp\frac{\C(\frac{1}{2}{-}n{-}iq)\hp\C(-\frac{1}{2}{-}n{-}iq)}
  {\C(-n{-}2iq)}\;,\label{norm}\\
  f_n(x)=[\xi(1{-}\xi)]^{-i\w_n}\hn
  F(-n,-n{-}2iq;{\ts\frac{1}{2}}{-}n{-}iq;\xi)\,.\label{fn}
\end{gather}
For the lowest few $n$, one can verify (\ref{norm}) by substituting~(\ref{fn}); in terms of $\xi$, the integral reduces to analytically continued Beta functions~\cite{Tam}.

Let us turn to the perturbation theory of (\ref{we-App}), in a form which assumes neither completeness of the QNMs nor finitely supported perturbing potentials. Since this only slightly generalizes previous analyses \cite{rmp,Tam}, the discussion will be brief. Let $V\mapsto V+\e V'+\e^2V''$ and $\w_n\mapsto\w_n+\e\w_n'+\e^2\w_n''$. The starting point is the derivative of (\ref{we-App}), $[d_x^2-V+\w_n^2]f_n'=[V'-2\w_n^\php\w_n']f_n$. The solution $f_n'$ is arbitrary up to two independent homogeneous solutions, say $f_n$ and $\chi_n$. The former, corresponding to an $\w$-dependent change in normalization, is irrelevant; the latter has to be adjusted so that $f_n'$ is outgoing. If the potentials decay at least exponentially and $\re\w_n\neq0$, imposing the boundary conditions does not involve any subtleties \cite{Alec,Wong}. Satisfying the \emph{two} conditions (for $x\To\pm\infty$) simultaneously is possible iff
\beq
  \w_n'=\frac{\lb f_nV'f_n\rb}{2\w_n\lb f_n^2\rb}\;;
\eeql{wn-pr}
note that the whole calculation is manifestly finite, and that neither the `matrix element' in the numerator nor the norm involves complex conjugation~\cite{rmp}. Analogously, one obtains in second order
\beq
  \w_n''=\frac{\lb f_n[(V'{-}2\w_n^\php\w_n')f_n'+V''f_n]\rb}{2\w_n\lb f_n^2\rb}
  -\frac{(\w_n')^2}{2\w_n}\;,
\eeql{wn-dpr}
which may be simpler than its counterpart in terms of logarithmic-derivative wave functions~\cite{Tam}. One can of course write $\lb f_n(V'{-}2\w_n^\php\w_n')f_n'\rb=\lb f_n'[d_x^2{-}V{+}\w_n^2]f_n'\rb$, but this does not seem to have an advantage. For $V$ even and $V'$ odd, the situation of the main text, $\w_n'=0$, but the outgoing condition on $f_n'$ remains non-trivial.

A few toy examples, in fact relevant to (\ref{Wel}), are
\begin{align}
  V_1(x,\e)&=\frac{V_0+\e}{\ch^{\hn2}\hn(x/2)}\quad\Rightarrow\quad
    V_1'=\ch^{\hn-2}\hn\Bigl(\frac{x}{2}\Bigr)\;,\quad\w_{1,n}'=\frac{1}{q}\;;
    \displaybreak[0]\\
  V_2(x,\e)&=\frac{V_0}{\ch^{\hn2}\hn[x/(2{+}\e)]}\quad\Rightarrow\quad
    V_2'=\frac{V_0}{2}\frac{x\sh(x/2)}{\ch^{\hn3}\hn(x/2)}\;,\notag\\[1mm] &
    \w_{2,n}'=\frac{1}{16q}+\frac{2n{+}1}{8}i\;;\displaybreak[0]\\
  V_3(x,\e)&=\frac{V_0}{\ch^{\hn2}\hn[(x{-}\e)/2]}\quad\Rightarrow\quad
    V_3'=\frac{V_0\sh(x/2)}{\ch^{\hn3}\hn(x/2)}\;,\notag\\[1mm] &
    \w_{3,n}'=0\;.
\end{align}
For the first few $n$,  $\w_{1(2),n}'$ can again be found explicitly from (\ref{wn-pr}), while $\w_{3,n}'=0$ is obvious since $f_n^2V_3'$ is odd. Working out these examples to the second order (\ref{wn-dpr}), and/or studying them asymptotically in~$n$, should be a useful (though not yet sufficient) stepping stone for the analogous analysis of (\ref{Uel})--(\ref{Ugr}).



\begin{thebibliography}{99}
\bibitem{odd}The $-\frac{4}{3}\e$ shift in $x(y)$ is not really needed for this, it just keeps the term as simple as possible.
\end{thebibliography}
\end{document}